# On the computational models for the analysis of illicit activities*

SARWAT NIZAMANI[a†], SAAD NIZAMANI[b], SEHRISH NIZAMANI[b] , IMDAD ALI ISMAILI[c]

[a]Department of Computer Science, Sindh University Campus Mirpurkhas
[b]Department of Information Technology, Sindh University Campus Mirpurkhas
[c]Pro-Vice Chancellor, Sindh University Campus Mirpurkhas

†sarwat@usindh.edu.pk

**Abstract-**This paper presents a study on the advancement of computational models for the analysis of illicit activities. Computational models are being adapted to address a number of social problems since the development of computers. Computational model are divided into three categories and discussed that how computational models can help in analyzing the illicit activities. The present study sheds a new light on the area of research that will aid to researchers in the field as well as the law and enforcement agencies.

*Part of this paper is taken from the Ph.D. thesis entitled "Computational models for analysis of illicit activities" by the first author.

## Introduction

There happen numerous illicit activities in our society, which, from time to time affect the people by harming individuals directly or indirectly. These illicit activities pose a threat to our society in some way or other. The illicit activities can be of many types, ranging from violating traffic laws which may hurt individuals, to burglary or organized terrorism threats (Chen et al., 2003). Researchers (Clause et al., 2007; Galam, 2002; Gutfraind, 2010; Nizamani et al., 2012; Sengupta el al., 2014) from distinguished disciplines have studied this sensitive problem in order to minimize its effect. In this regard, computational models play an important role to assist law enforcement agents in order to curb certain illicit acts to some extent. The recent study (Sengupta el al., 2014) shows that the use of computer-aided technology in crime analysis has declined the crime rate in United States. Researchers from the field of computer science often use the techniques from data mining, social network analysis, natural language processing, etc., for the analysis of the criminal activities.

In this paper, we present the study of computational models that have been developed for the analysis of illicit activities. The research for analyzing the illicit activities has been conducted by the researchers from distinguished domains. Scholars have used different research perspectives, employed diverse methods and developed various tools. For instance, from the computing science's perspectives, the crime data mining (Xu and Chen, 2005) and investigative data mining (Memon et al. 2007) approaches have been widely used. Both disciplines essentially use the techniques from "data mining" for discovering the patterns from available sources for the purpose of investigating crimes and help law enforcement agencies, in order to strategize against crimes. We have divided the study into three categories of computational models for the analysis of illicit activities. These categories include: criminal network analysis; text analysis; and mathematical models for analysis of illicit activities as shown in Figure 1. Each of these categories is discussed in the following sections.

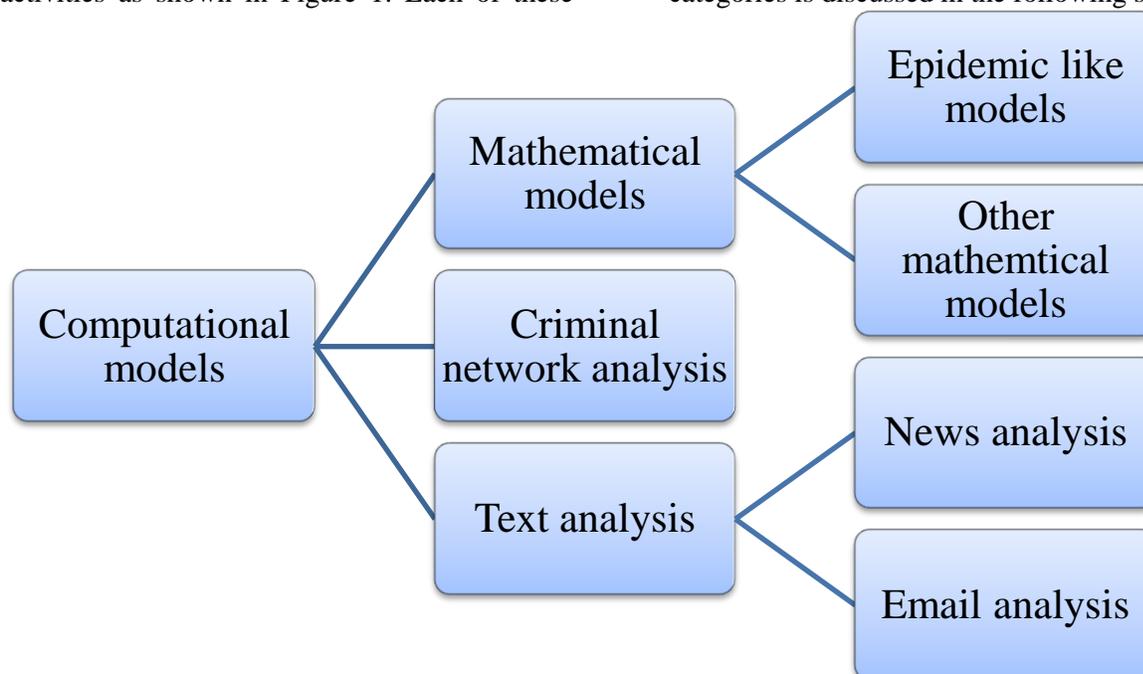

## Material and methods
### Criminal network analysis

Criminal network analysis is considered as an essential step for analysis of the organized criminal activities. The authors (Xu and Chen, 2005) presented a framework which is named as CrimeNet Explorer for analysis of the criminal networks. The underlying framework is composed of four phases i.e. (i) network creation; (ii) network partition; (iii) structural analysis; and (iv) network visualization. The network creation phase constructs the criminal network from the criminal-justice data which is collected from the crime incident reports by using concept space approach. The entities and the relations among the entities are extracted, which result in the form of criminal network. Via a hierarchical clustering approach, the underlying network is then partitioned in order to reveal sub-groups in the network. In the next stage, the structural analysis of the network is performed to discover the key actors in the network. At the last stage, network visualization details the network, graphically, which is informative too.

In another study, Chen et al. (2003) described four case studies of crime data mining techniques, which the authors have done in their project COPLINK. The case studies include (i) entity extraction from police narrative reports; (ii) detecting criminal identity deceptions with an algorithmic approach; (iii) authorship analysis in cybercrimes; and (iv) criminal network analysis. In the entity extraction case study, the authors extracted named entities from the criminal justice free text (police narrative reports). The authors employed neural network (Gershenson, 2003) based entity extraction, which extracts useful entities from police narrative reports. Criminal identity deception case study uses an algorithmic approach for detecting deception from criminal names, addresses, social security numbers and date of birth. The authors used Euclidean distance-based deception detection, which identifies the real identities by computing the similarities between real identity and the identities given by the criminals. Authorship deception is a common illicit activity in the cyber-world and the task of authorship deception detection is called authorship analysis. For the case study of the authorship analysis, the authors used the support vector machine (SVM) (Vapnik, 2000) classification method. For the criminal network analysis, the authors employed social network analysis (SNA) techniques. SNA techniques are used to discover hidden patterns from the criminal networks, which include network extraction, sub-group detection, key actor identification and discovering interaction patterns in the network.

A study by Allanach et al. (2004) proposed a Hidden Markov Model (HMM)-based terrorist network detection, tracking and counteracting model. Each terrorist activity is considered to be a state which is treated as a transaction to be carried by terrorists in the network.

The evolution of a network discloses the origin of the network and its construction patterns. To explore its origins, a hierarchical-based terrorist network evolution method is proposed by Nizamani and Memon (2011), which produces a simplified network evolution model. In this way, one can comprehensively visualize that how the network started to evolve and who are the key members of the network (as well as how they are grouped)? In the hierarchical evolution, the analyst can examine distinct groups at various levels of the hierarchy. Furthermore, in the study, various measures of cluster quality have been computed that show the goodness of hierarchical communities in the network. The evolution proceeds by grouping of members from single member groups and continues until forming one large group including all members.

## Text analysis for knowledge discovery about illicit activities

In the literature, there are a number of articles that use text analysis techniques involving text classification methods for analysis of illicit activities. For instance, suspicious email detection (Nizamani et al., 2012; Appavu et al., 2007), email authorship analysis (De vel et al. 2001; Iqbal et al., 2008; Iqbal et al., 2013; Zheng et al., 2006 ; Nizamani and Memon, 20013), criminal identity deception (Chen et al. 2004) and identification of criminal entities from the text(Chen et al. 2004) use text analysis techniques. The study (Jiang and Tan, 2010) presents an ontology based terrorism concept extraction from the web. A domain lexicon is constructed manually and the input text is given to the concept extraction system, which uses the algorithms to extract important concepts from the text based on domain ontology. A study by

Nizamani et al. (2011) presented a cluster-based text classification model for the task of suspicious email detection. In another study (Nizamani and Memon, 2013), the same model is applied for the task of email authorship identification.

Text analysis study has also been conducted for the news analysis, which supplements analysts' information for analyzing criminal activities. The development of the Internet sources has empowered the news channels to the enormous amount of news. While the volume of information is immense, if it is not utilized, it is merely a waste of cyber-space. The news can be analyzed to glean useful, specific knowledge and identify patterns. For example, in a study (Krstajic et al.,2010 ), a news analysis method is proposed which identifies entities in the news and extracts spatio-temporal information about those entities. The visual analysis of the entities shows that how frequently certain entities appeared in the news articles in a time-window. Furthermore, the study extracts the co-occurrences of the entities in different news articles, which determines the relationships among the entities.

The study of the text analysis for analyzing the news has also been conducted in the languages other than English. For instance, the study (Eldin and El-Beltagy, 2013) presents an auto-tagging of Arabic news using Wikipedia. The approach used by the authors is comprised of two phases. The first phase builds the concept dictionary, which is composed of four steps, namely; (i) extraction of required information from Wikipedia; (ii) filtration and partitioning of entries; (iii) dictionary normalization; and (iv) building of inverted index. In the second phase, the input text is tagged. This includes the steps such as, (i) sentence boundary identification; (ii) word tokenization; (iii) extraction of candidate tag anchor; and (iv) matching of candidate tags to the concept entries in the dictionary.

The study of the news analysis presented by Nizamani and Memon (2012) dissects the news for the purpose of in-depth analysis. The analyzed news are presented in such a way that one can give a comprehensive look at the news. The main theme of the news and entities involved can be highlighted with spatial and temporal information of the news. The study has also been extended to analyze the criminal activities and criminal convictions in areas of interest from FBI press releases, using techniques from data mining.

## Mathematical models for the analysis of illicit activities

The models behind knowledge discovery of the criminal networks, analysis of the text to discover knowledge about illicit activities or quantitative analysis of the illicit activities involve mathematics in the back end. Mathematics has immense importance within the realm of national security (Memon et al., 2009). Mathematical and computational models have great power for analyzing the illicit activities. Social science researchers employ computational models for investigating crimes and criminal behavior (Brantingham et al., 2009).

The mathematical models that are often employed in the study of illicit activities generally involve graph theory for mapping individuals (criminals) and other entities as nodes and relationship (links/ edges) among them. A number of properties are then often extracted, which disclose the key criminals and the role or position of individuals in conducting the crime. The studies (Lindelauf et al., 2008; Maeno and Ohsawa, 2009; Memon et al. 2007) employed the graph theory for the analysis of covert networks behind the terrorist attacks.

Game Theory (Sandler and Daniel, 2003) is often applied by the mathematicians and has wide applications in counterterrorism. Terrorists and government are considered to be the players of the game, planning strategies against each other (Sandler and Daniel, 2007). The study (Melese, 2009) proposes a brinkmanship game theory model in which the key players are United Nations and terrorists. The United Nations is considered as the "principal" and terrorists are referred to as "agents". In the model, the principal issues a brinkmanship threat, i.e. preemptively acquiring the weapons of mass destruction (WMD) from the terrorists (Melese, 2009). The study describes the conditions for the credibility of the brinkmanship threat, when it is suitable to issue the brinkmanship threat and when it may imperil the world community.

Physicists have also contributed to the area. Galam (2002) well known socio-physicist (also known as father of Socio-Physics) modeled the September 11 events as a global threat of terrorism. Galam discussed the role of passive supporters, that how they put the entire world under the risk of terrorism. Galam modeled the whole world as a square lattice and considered each cell of the lattice as being occupied by a single individual of the world. Galam emphasized on the role of passive supporters distributed at varying locations in the world and computed the proportion of the passive supporters, who can make the entire world vulnerable to terrorism. The concept of percolation (Shante and Kirkpatrick, 1971) is used in the study and a threshold is calculated to describe the proportion of passive supporters that can cause such risk. Galam in same article argues that with a certain coverage of passive supporters (i.e. when the number of passive supporters exceeds the percolation threshold), no region in the world would be safer from the global terrorism risk and current military operations would not be sufficient to curb the global terrorism. In another article Galam and Mauger (2003) studied the global terrorism with respect to passive supporters. The authors suggest that by modifying the percolation threshold in virtual space, the global terrorism threat can be reduced to local geographic problem. The authors considered every individual of the global population and the world is regarded as a multidimensional virtual space. The authors used following formula for calculating the percolation threshold $p_c$.

$$p_c = a[(d-1)(q-1)]^{-b}$$

Where d is the space dimensionality and q is the local connectivity; while a = 1.2868 and b = 0.616. The equation shows that $p_c$ is a decreasing function of q connectivity and d dimensionality. The authors reveal that with d =5 and q=20, even with the 10% of passive supporters the entire world would be at risk.

Beside these definite criminal activities, there happen some violent events which are caused by some critical shocks. The study (Roehner, 2007) presents a response function to such critical shocks in the social sciences. Specifically, the authors analyzed the aftermath of September 11 and the destruction of the Babri Mosque in 1992 in India. The study showed the relationship between the number of news articles published on those events in relation to the time of the occurrence of the events. Furthermore, the study analyzed the financial market response on those events.

The epidemic-like model of the critical shocks is presented by Nizamani et al. (2014) which analyses the specific world issues that lead to violence. The model illustrates the dynamics of the population which is sensitive to those shocks. The model is based on differential equations and is comprised of five types of agents belonging to that population group. The model demonstrates the way initially very few upset agents cause dynamics in the overall population. The epidemic like behavior of the model shows the rise of the public outrage, leading to the public violence and finally ending the dynamics. In the study, the outbreak conditions have been determined to whether consider the issue as serious or not.

The highlights of a few of the computational models discussed above are illustrated in Table 1, along with category of the model, methodology, title and the reference.

Table 1: Highlights of the computational models for analysis of illicit activities

| Category | Methodology | Title | Paper |
|---|---|---|---|
| Criminal network analysis and text analysis (Police narrative reports) | Data mining | Crime data mining: an overview and case studies | Chen et al.(2003) |
| Criminal network analysis and text analysis (Police narrative reports) | Data mining | Crimenet explorer: a framework for criminal network knowledge discovery | Chen et al. (2005) |
| Criminal network analysis | Investigative data mining | How investigative data mining can help intelligence agencies to discover dependence of nodes in terrorist networks. | Memon et al.(2007) |
| Criminal network evolution | Data Mining | Evolution of Terrorist Network Using Clustered Approach: A Case Study | Nizamani and Memon (2011) |
| Criminal network analysis | Statistical model HMM | Detecting, tracking, and counteracting terrorist networks via hidden markov models | Allanach et al.(2004) |
| Text analysis(Suspicious email detection) | Data mining/machine learning | . Modeling suspicious email detection using enhanced feature selection | Nizamani et al. (2012) |
| Text analysis(Suspicious email detection) | Data mining/machine learning | Association rule mining for suspicious email detection: A data | Appavu et al.(2007) |

| | | mining approach | |
|---|---|---|---|
| Text analysis(Spam email detection) | Data mining/ clustering | Spam detection using text clustering | Minoru and Shinnou (2005) |
| Mathematical modeling | Percolation theory | Passive support of terrorism | Galam(2002) |
| Mathematical modeling/ Models from physics | Percolation theory | On reducing power of terrorism | Galam and Mauger (2003) |
| Epidemic like model of public outrage | Mean field equations | From public outrage to the burst of public violence: An epidemic-like mode | Nizamani et al.(2014) |

## Conclusions

In this paper we presented the study on the computational models for the analysis of illicit activities. The study analyzed the computational model from three perspectives; namely; criminal network analysis; text analysis; and mathematical models. The study discusses the advancements in the field by elaborating the current research trends in the field.